\documentstyle[multicol,aps,epsf]{revtex}
\newcommand{\beq}{\begin{equation}}
\newcommand{\eeq}{\end{equation}}
\begin{document}
\tightenlines

\title{Fermi-condensate quantum phase transition
in high temperature superconductors}

\author{V.R. Shaginyan
\footnote{E--mail: vrshag@thd.pnpi.spb.ru}}
\address{
Petersburg Nuclear Physics Institute, Gatchina 188350, Russia}
\maketitle

\begin{abstract}
The effect of a quantum phase transition associated with the
appearance of the fermion condensate in an electron liquid on the
properties of superconductors is considered. It is shown that the
electron system in both superconducting and normal states exhibits
characteristic features of a quantum protectorate after the point of
this fermion-condensation quantum phase transition. The
single-particle spectrum of a superconductor can be represented by
two straight lines corresponding to two effective masses $M^*_{FC}$
and $M^*_{L}$. The $M^*_{FC}$ mass characterizes the spectrum up to
the binding energy $E_0$ and $M^*_L$ determines the spectrum
at higher binding energies. Both effective masses are retained in the
normal state. These results are used to explain the lineshape of
single-particle excitations and other properties of
high-$T_c$ superconductors and are in a good agreement with recent
experimental data.
\end{abstract}

{\it Keywords}: {\small Strongly correlated electrons;
Superconductivity}

\begin{multicols}{2}

Recent experiments gave rather accurate data on the dispersion of
single-particle excitations over a wide range of binding energies
\cite{vall}. These experiments were carried out with high-$T_c$
superconductors Bi$_2$Sr$_2$CaCu$_2$O$_{8+\delta}$ differing in the
doping level both at temperatures $T$ below the critical temperature
$T_c$ and at $T_c\leq T$. It was inferred that the dispersion
of quasiparticle excitations can be described
in the energy range (--200--0) meV by two straight lines intersecting
at the binding energy $E_0\simeq(50 - 70)$ meV \cite{vall}. This
circumstance directly points to the existence of a new energy scale
in the self-energy part of quasiparticle excitations at temperatures
$T\leq T_c$ and $T_c\leq T$ \cite{vall}. Therefore, new additional
constraints can be imposed on the theories that are in principle
applicable to the description of properties of high-$T_c$
superconductors. Experimental data on single-particle electron
spectra of high-$T_c$ superconductors indicate
that the perturbation of the superconducting phase and
single-particle spectra by phonons, collective states, or impurities
is very small. Therefore, this state can be described as a strongly
collectivized quantum state or as ``quantum protectorate''
\cite{rlp}. At $T=0$, the
ground state energy $E_{gs}[\kappa({\bf p}),n({\bf p})]$
of two-dimensional electron liquid
is a functional of the order parameter of
the superconducting state $\kappa({\bf p})$ and occupation numbers
$n({\bf p})$ and is determined by the known equation of the
weak-coupling theory of superconductivity
\beq E_{gs}=E[n({\bf p})]
+\int V({\bf p}_1,{\bf p}_2)\kappa({\bf p}_1)
\kappa^*({\bf p}_2)
\frac{d{\bf p}_1d{\bf p}_2}{(2\pi)^4}.\eeq
Here  $E[n({\bf p})]$ is the ground-state energy of the normal
Fermi liquid, the pairing interaction $V({\bf p}_1,{\bf
p}_2)$ is assumed to be weak, while
$n({\bf p})=v^2({\bf p})$, and
$\kappa({\bf p})=v({\bf p})\sqrt{1-v^2({\bf p})}$.
Minimizing $E_{gs}$ with
respect to $\kappa({\bf p})$ we obtain the equation for
the superconducting gap $\Delta({\bf p})$
\beq \varepsilon-\mu=\Delta
\frac{1-2v^2({\bf p})} {2\kappa({\bf p})};
\Delta=-\int
V({\bf p}, {\bf p}_1) \kappa({\bf p}_1)
\frac{d{\bf p}_1}{4\pi^2}.\eeq
where the single-particle energy $\varepsilon({\bf p})$ is
determined by the Landau equation,
$\varepsilon({\bf p})=\delta
E[n({\bf p})]/\delta n({\bf p}),$
$\mu$ is chemical potential.
If $V\to 0$, then, the gap $\Delta({\bf p})\equiv
0$, and Eq. (2) is reduced to the equation proposed in
\cite{msh} \beq \varepsilon({\bf
p})-\mu=0,\: {\mathrm {if}}\,\,\, 0<n({\bf p})<1;\: p_i\leq p\leq
p_f.\eeq
This equation defines a Fermi liquid of a new type \cite{vol} for
which the order parameter $\kappa({\bf p})$ differs from zero in the
$L_{FC}$  range
of momenta $p_i\leq p\leq p_f$; the occupation numbers
$n({\bf p})=1,0$ outside the $L_{FC}$ range, as it must be in the
normal Fermi liquid.
It follows from Eq. (4), that a  fermion
system with the fermion condensate (FC) is broken into two
quasiparticle subsystems: the first subsystem in the $L_{FC}$ range is
occupied by the quasiparticles of the effective mass $M^*_{FC}\to
\infty$, while the second one is occupied by quasiparticles of finite
mass $M^*_L$ with momenta $p<p_i$. The FC appears in an electron system
at a low density, when the effective electron-electron interaction
constant is sufficiently large. In a common electron liquid, this
constant is directly proportional to the dimensionless parameter
$r_s\sim1/p_Fa_B$, where $a_B$ is the Bohr radius and $p_F$ is the
Fermi momentum.  The FC occurs at a certain $r_s=r_{FC}<r_{cdw}$ and
precedes the appearance of a charge-density wave (CDW) at
$r_s=r_{cdw}$.  This phase transition occurs at $T=0$ when the
parameter $r_s$ attains its critical value $r_{FC}$ and represents a
quantum phase transition.  At $r_s>r_{FC}$ and $r_s-r_{FC}\ll r_{FC}$,
the region $p_f-p_i$ occupied by the Fermi condensate is
$(p_f-p_i)/p_F\sim r_s-r_{FC}$.  Because the order parameter of the
fermion-condensation quantum phase transition (FCQPT) is $\kappa({\bf
p})$, the maximum value of the gap $\Delta_1\sim V$, as it follows from
Eq. (2), while $\kappa({\bf p})$ is determined by the relatively strong
particle-hole interaction.  Therefore, the perturbation of the
parameter $\kappa({\bf p})$ by $V$ can be neglected in the first
approximation. At $T=0$, $M^*_{FC}$, $E_0$, and $\Delta_1$ can be
calculated from Eq. (2) \cite{msh} \beq \frac{p_F}{M^*_{FC}}
\simeq\frac{2\Delta_1}{p_f-p_i};\,\,\,
E_0\simeq2\frac{(p_f-p_F)p_F}{M^*_{FC}}\simeq
2\Delta_1;\eeq
$$\Delta_1\simeq 2\frac{\beta\varepsilon_F(p_f-p_F)}{p_F}.
$$
Thus, electron system with the FC is
characterized by two finite effective masses $M^*_{FC}$ and $M^*_L$,
the single-particle dispersion at $p\sim p_F$ can be
approximated by two straight lines,
$E_0$ does not depend on $p_f-p_i$,
and $M^*_{FC}$ is proportional to this difference.
Taking the usual values of the dimensionless coupling constant
$\beta\simeq VM^*_L/(2\pi)\simeq 0.3$, $(p_f-p_F)/p_F\simeq 0.2$, we
obtain from Eq.  (4) that $\Delta_1\simeq 0.1\varepsilon_F$, with
$\varepsilon_F$ being the Fermi energy. Because $T_c\simeq \Delta_1/2$,
we conclude that $T_c$ reaches high values. At finite $T$, Eq. (4) is
replaced by \cite{msh}
\beq \frac{p_F}{M^*_{FC}}\simeq \frac{4T}{p_f-p_i};\,\,\,
E_0\simeq\frac{(p_f-p_i)p_F}{M^*_{FC}}\simeq 4T. \eeq
By comparing Eqs. (4) and (5), we see that $M^*_{FC}$ and
$E_0$ weakly depend on temperature at $T\leq T_c$.
If there exists the pseudogap above $T_c$, then $T_c$ is to be
replaced by $T^*$ because in that case $2\Delta_1\simeq T^*$
\cite{msh}. The above consideration shows that the form of the
single-particle spectrum $\varepsilon({\bf p})$ is determined by the
FCQPT and, therefore, its form is universal being almost independent of
the contribution coming from impurities, phonons, etc. Actually, the
particle-hole interaction defines only the region $L_{FC}$ occupied
by the condensate after the point of the FCQPT.
Thus, a system with the FC is characterized
by a universal form of the single-particle spectrum and possesses
quantum protectorate features at $T\ll T_f$, with $T_f$ being a
temperature at which the effect of the FCQPT disappears \cite{msh}.
Now turn to the description of the experimental data
\cite{vall}. Experimental studies showed that the Van Hove
singularity is located at the $(\pi,0)$ point of the
Brillouin zone, and an almost dispersionless section of the spectrum
is observed in this region. This
allows the suggestion to be made that the FC is arranged about
this point. The line $Y\Gamma$, which is
known as the line of zeros of the Brillouin zone, passes through the
points $(\pi,\pi)$ --- $(0,0)$ at an angle of $\pi/4$ to the line $Y
\bar{M}$ passing through the points $(\pi,\pi)$ --- $(\pi,0)$.
Single-particle spectra
were measured along the lines parallel to $Y\Gamma$ and
$Y\bar{M}$, from the line of zeros to the $Y\bar{M}$ line.
It was shown that $E_0$ is constant for a given sample being
independent of the angle $\phi$, reckoned from $Y\Gamma$ to
$Y\bar{M}$. The angle (kink) between the straight line
characterizing the single-particle spectrum grows with increasing
$\phi$ and with decreasing doping level. This general
pattern is retained at $T>T_c$ \cite{vall}. To describe these
experimental data, we use the following results:
$r_s$ grows with decreasing
the doping level and exceeds the critical value $r_{FC}$ in the
overdoping region; the difference $(p_f(r_s)-p_i(r_s))$ as a function
of $r_s$ grows with increasing $r_s$; the difference
$(p_f(\phi)-p_i(\phi))$ increases with increasing $\phi$ and attains a
maximum at $(\pi,0)$. Noticing
that strong fluctuations of the charge density or CDW are observed in
undoped samples, we may conclude that the formation of the FC in copper
oxides is a quite determinate process stemming from the general
properties of a low-density electron liquid. Then, Eq. (4) shows that
$E_0$ does not depend on $\phi$ and $T$ at $T\leq T_c$, while the kink
increases with increasing $\phi$, because $M^*_{FC}$ linearly depends
on $(p_f(\phi)-p_i(\phi))$. Then, $E_0\simeq (50 - 70)$ meV
\cite{vall}, which is in agreement with Eq. (4), because $E_0\simeq
2\Delta_1$ in these materials. Then, the dispersion kink must
increase with decreasing doping level. As it follows from Eq. (4),
$E_0\simeq 2\Delta_1\sim(p_f-p_F)$, the kink point shifts towards
higher binding energies as the doping level decreases.
Consider the lineshape $L(q,\omega)$ being a function of two variables.
Measurements carried out at a fixed binding energy $\omega=\omega_0$,
where $\omega_0$ is the energy of a single-particle excitation,
determine the lineshape $L(q,\omega=\omega_0)$ as a function of
momentum $q$.  Eqs. (4) and (5) show that $M^*_{FC}$ is finite and the
system behaves as a normal liquid at the energies $\omega<2\Delta_1$,
while the width of single-particle excitations is of the order of $T$
\cite{msh}. This behavior was observed in experiments on measuring the
lineshape at a fixed energy \cite{vall}. The lineshape can also be
determined as a function $L(q=q_0,\omega)$ at a fixed
$q=q_0$.  At small $\omega$, $L(q=q_0,\omega)$ has a characteristic
maximum and width. At energies $\omega\geq 2\Delta_1$, quasiparticles
of mass $M^*_{L}$ come into play, leading to a growth of the function.
One may also use the Kramers-Kr\"{o}nig transformation to construct the
imaginary part of the self-energy starting with the real one. As a
result, the lineshape $L(q=q_0,\omega)$ possesses the known
peak-dip-hump structure directly defined by the existence of the two
effective masses $M^*_{FC}$ and $M^*_L$.
All these results are in good qualitative agreement with the
experimental data \cite{vall}.  This work was supported by the Russian
Foundation for Basic Research, project no.  01-02-17189.

\end{multicols}

\end{document}